\documentclass[%
 reprint,
 amsmath,amssymb,
 aps,
]{revtex4-2}
\usepackage{graphicx}
\usepackage{dcolumn}
\usepackage{bm}
\usepackage{braket}
\usepackage{siunitx}
\usepackage{upgreek}
\usepackage{fontspec}
\begin{document}

\preprint{APS/123-QED}

\title{Demonstration Of A Quantum Magnetometer Chip Based On Proprietary And Scalable 4H-Silicon Carbide Technology}

\author{P. A. Stuermer$^{1,3}$, D. Wirtitsch$^{2}$, T. Steidl$^{3}$, R. Wörnle$^{3}$,  J. Körber$^{3}$, W. Schustereder$^{1}$, C. Zmoelnig$^{1}$, P. Urlesberger$^{1}$, F. Chiapolino$^{1}$, S. Meinardi$^{1}$, K. Edelmann$^{4}$, M. Kern$^{5}$, J. Anders$^{4,5}$, S. Krainer$^{1}$, H. Heiss$^{6}$, M. Trupke$^{2}$ and J. Wrachtrup$^{3}$}
 
\affiliation{$^{1}$Infineon Technologies Austria AG, Villach, Austria \\
             $^{2}$Austrian Academy of Sciences, Vienna, Austria\\
             $^{3}$3rd Institute of Physics, University of Stuttgart, Germany\\
             $^{4}$Institut für Mikroelektronik Stuttgart (IMS CHIPS), Stuttgart, Germany\\
             $^{5}$Institute of Smart Sensors, University of Stuttgart, Germany\\
             $^{6}$Infineon AG, Munich, Germany \\}

\

\date{\today}

\begin{abstract}
This work presents an industrially scalable, power‐efficient and high‐performance quantum magnetometer chip based on proprietary 4H‐silicon carbide (SiC) technology, leveraging wafer‐scale fabrication techniques to optimize V2 silicon vacancy color centers for highly reproducible, industry‐grade fabrication with precise control of depth and density. The integration of these color center ensembles into a planar silicon carbide waveguide enables efficient excitation of a large ensemble and simplifies fluorescence extraction compared to standard confocal methods. We report continuous-wave (CW) optically detected magnetic resonance measurements, complemented by Rabi, Ramsey, and Hahn-echo sequences, which demonstrate coherent capabilities of the large embedded ensemble of V2 centers. Based on the data, our device exhibits sensor shot-noise limited sensitivities 2-3 orders of magnitude lower compared to more complex confocal techniques. Collectively, these advancements simplify the quantum sensor architecture, enhance sensitivity, and streamline optical excitation and collection, thereby paving the way for the development of next‐generation SiC-quantum sensing technologies.
\end{abstract}

\maketitle


\begin{figure*}[ht]
  \centering
  \includegraphics[width=\textwidth]{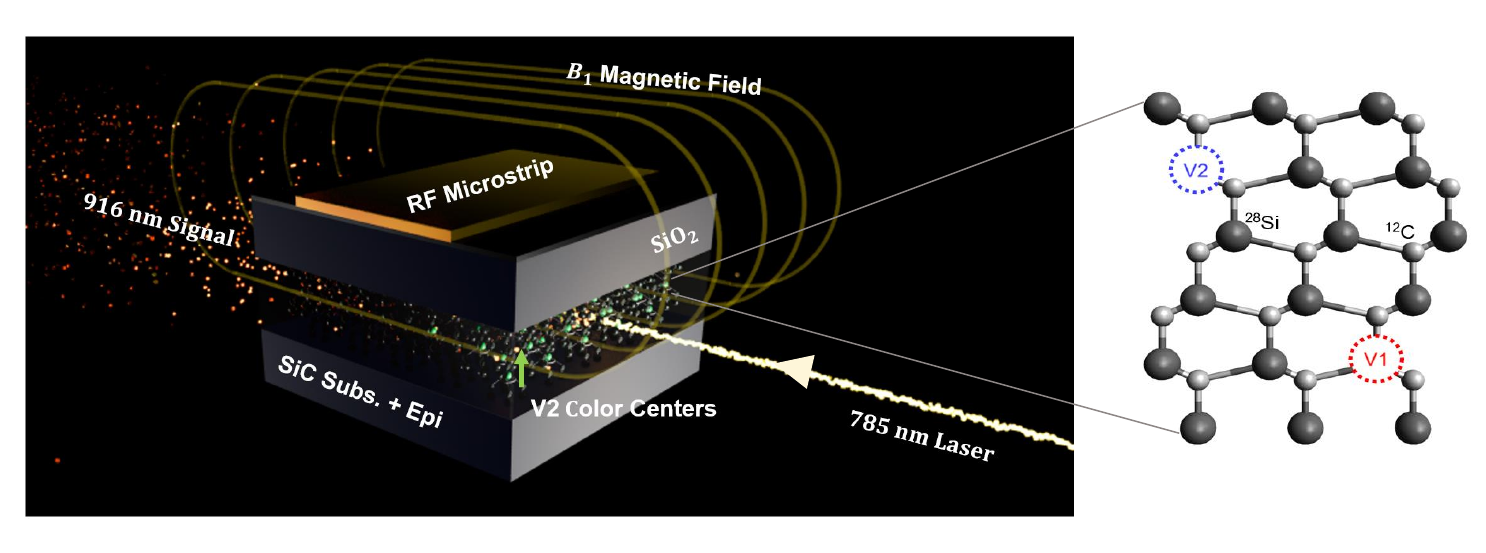}
  \caption{Illustration of the sensor concept: The planar monolithic 4H-SiC waveguide is formed as sandwich structure consisting of a n\textsubscript{++} doped substrate, an intrinsically doped core and a very thin layer of SiO\textsubscript{2} on top. The V2 C3\textsubscript{v}-symmetry axis points perpendicular to the sensor plane, enabling efficient RF-excitation by a simple microstrip design as shown above or alternatively by a coil. On the right, a schematic of the 4H-SiC crystal structure is shown, with corresponding silicon vacancy color centers, V1 and V2.}
  \label{fig:Figure1}
\end{figure*}
\section{Introduction}
\begin{figure*}[ht]
  \centering
  \includegraphics[width=\textwidth]{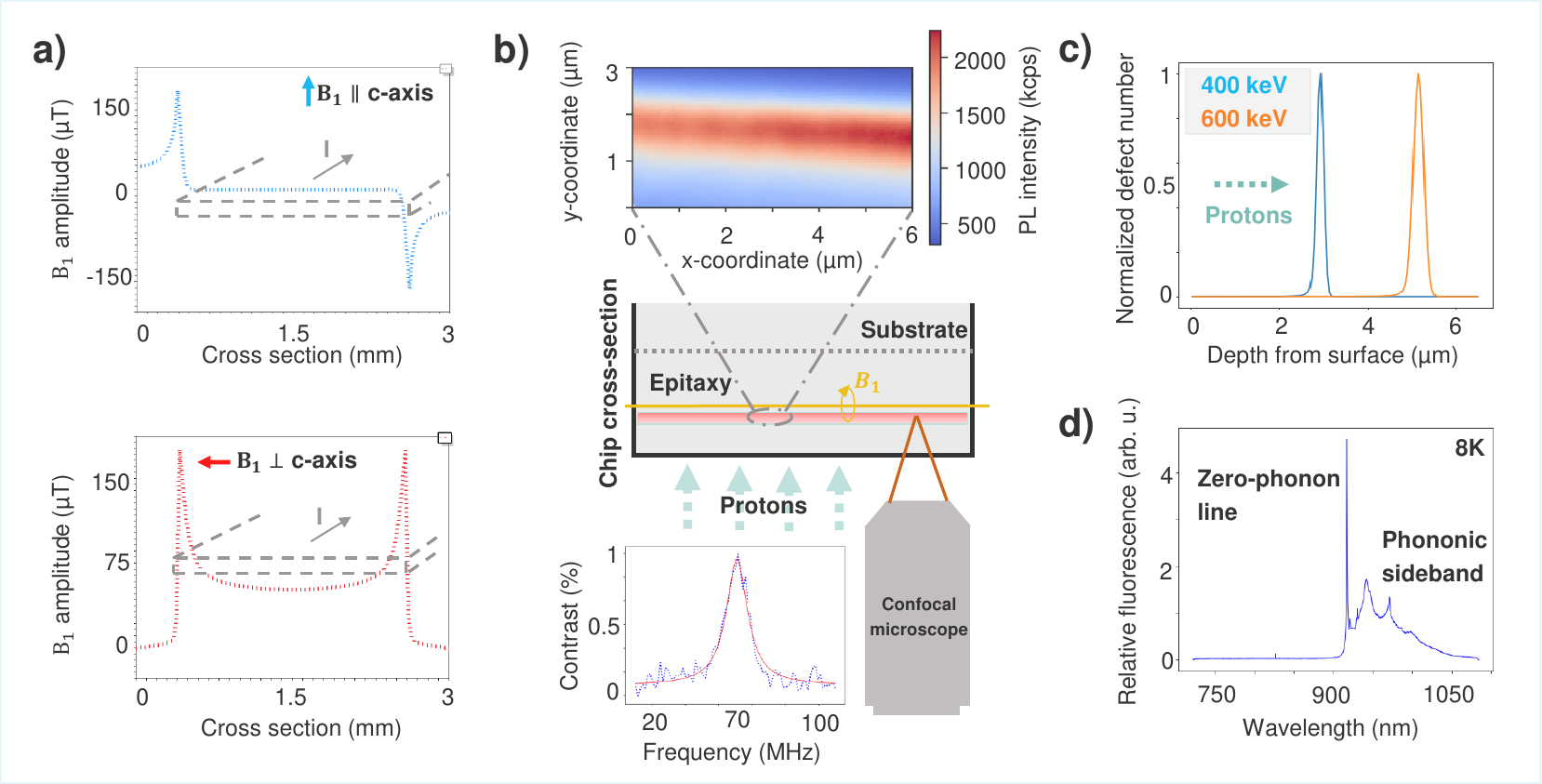}
  \caption{a) Calculated $B_{1}$ field distributions induced by a 70 MHz current parallel and orthogonal to the color center symmetry axis at a distance of \SI{3}{\micro m} from the microstrip.  b) Confocal scan of the chip cross section, revealing a ribbon of V2 emerging from the proton implantation peak and a corresponding CW optical detected magnetic resonance (ODMR) measurement, fitted with a Lorentzian. c) Simulated defect density distribution induced by protons implanted into 4H-SiC with 400 keV and 600 keV. d) 8 K low-temperature spectrum of the V2 ensemble clearly showing the zero-phonon line (ZPL) at 916 nm and its phononic sideband.}
  \label{fig:Figure2}
\end{figure*}

Spin-based quantum sensing based on NV-centers in diamond \cite{castelletto2023quantum} has been developed for more than 10 years and already found its way into commercialization in form of a multitude of companies and startups employing its 
physical principles in areas ranging from biotech to material analysis \cite{yole2023quantum}. However, there are major 
obstacles for diamond-based quantum technologies. Besides being a much more expensive material than 
silicon carbide, it does not fare well within the semiconductor manufacturing ecosystem as it cannot 
leverage existing scalable wafer-based processes that come with high-volume SiC fabrication \cite{castelletto2023quantum, sekiguchi2024diamond, barry2020sensitivity, wirtitsch2023exploiting, gruber1997scanning}. Silicon carbide has a wide bandgap and hosts stable crystal defects that give rise to solid-state spin quantum systems with advantageous quantum properties \cite{simin2016all, widmann2015coherent, widmann2018bright, widmann2019electrical, steidl2024single, nagy2019high, heiler2024spectral}. In this work, the negatively charged V2 color center is used \cite{niethammer2016vector, niethammer2019coherent}, which is formed by the absence of a silicon atom within a cubic lattice site and a subsequent capture of a single electron. Its crystal structure is shown in Figure \ref{fig:Figure1}. The system can be optically excited off-resonantly via 785 nm, resulting in a zero-phonon line at a wavelength of \SI{916}{nm} and a phononic sideband extending up to 1100 nm \cite{soykal2016silicon}. Even though silicon carbide is a younger quantum 
technology platform compared to diamond, recent demonstrations of novel defect systems show quantum properties 
on par with diamond, making SiC a very promising candidate for high-volume quantum devices \cite{li2022room}. 
Quantum sensing can be implemented for different sensing parameters such as magnetic and electric fields, 
temperature, pressure and rotation \cite{abraham2021nanotesla, likhachev2025all, gottscholl2024enhancing, gottscholl2024operation, soshenko2021nuclear, kraus2014magnetic}, each requiring slightly different implementations but fundamentally relying on the same physical principle: detection of spin-dependent transitions. Here, in order to achieve high sensitivity, one requires a strong fluorescence signal contrast, narrow linewidth and a large number of fluorescence photons \cite{barry2020sensitivity, simin2017quantum, lekavicius2023magnetometry}. One method in order to increase the number of fluorescence photons is to expand
ensemble density. However, higher ensemble densities lead to higher crystal damage, in turn causing severe deterioration of quantum properties leading for example to increased linewidth and reduced contrast. Another strategy is to increase the active volume while keeping the ensemble density constant. However, this requires an optical waveguide \cite{babin2022fabrication, bosma2022broadband, radulaski2017scalable, krumrein2024precise} in order to efficiently excite the color centers and extract 
their signal, which poses a technological challenge on its own. In absence of a photonic waveguide, the required optical excitation power over a large volume increases, affecting the overall energy consumption and therefore competitiveness of the sensor-system. Furthermore, the unconfined emission of photons reduces collection efficiency, in turn leading to lower sensitivity. \\
\\
In order to circumvent those problems, our sensor incorporates a monolithic SiC waveguide as shown in Figure \ref{fig:Figure1} a). This photonic concept \cite{stuermer2025low} is based on exploiting the change of the refractive index of SiC when changing the doping level of the semiconductor. By using a n\textsuperscript{++} doped substrate as bottom cladding, a lowering of the refractive index is achieved. On top, a very thin layer of SiO\textsubscript{2} is deposited, causing a strong change of the refractive index compared to the the intrinsically doped core, leading to asymmetric modes along the vertical stack. The number of waveguide modes is dependent on the thickness of the core layer \cite{stuermer2025low}. Additionally, the strong jump in the refractive index between the core and upper layer additionally has the effect of strongly confining the evanescent portion of the wave propagating in the upper cladding, avoiding coupling to conductive structures on top of the SiO\textsubscript{2}. The photonic waveguide shows near-identical behavior for wavelengths from around 780 nm up to 1200 nm, guiding both the excitation and the fluorescence light of the V2 color centers efficiently with a low loss \cite{stuermer2025low}. \\ 
\\
Besides efficient photonic excitation and collection, a V2 spin-based quantum sensor additionally needs a radio-frequency (RF) field, termed $B$\textsubscript{1}, to drive spin-transitions \cite{carter2015spin, weber2011defects, nagy2018quantum}. The strength of the excitation of the spin-system in first-order perturbation is given by \cite{simin2017quantum}:
\begin{equation}
    P = \bra{\Psi_\textsubscript{b}} B_\textsubscript{1} S \ket{\Psi_\textsubscript{a}}
\end{equation}
where $B$\textsubscript{1} is the amplitude vector of the driving RF field and $S$ is the tensor of spin matrices for the silicon vacancy spin system, while $\Psi_a$ and $\Psi_b$ are the corresponding quantum states involved in the transition. This matrix element is maximized for the case of $B$\textsubscript{1} orthogonal to the spin-system symmetry axis \cite{simin2017quantum}. Accordingly, in our design the crystal axis and therefore the C3\textsubscript{V}-symmetry axis of the silicon vacancies points perpendicular to the sensor plane, simplifying the task to design a RF-microstrip with high $B$\textsubscript{1} homogeneity along nearly the whole chip area. One possibility is to deposit a metallization with thickness of \SI{3}{\micro m} on top of the SiO\textsubscript{2}, which generates a homogeneous field parallel to the sensor plane, when current is applied. The $B$\textsubscript{1} fields at the center of the microstrip are calculated via Ansys finite-element simulations (FEM) for an exemplary current of 200 mA at a frequency of 70 MHz at distance of \SI{3}{\micro m} with respect to the microstrip and results are shown in Figure \ref{fig:Figure2} a). We note that efficient coils within the range of V2 frequencies around the zero field splitting can be designed for excitation as well. Those have the advantage of lower current consumption for same $B$\textsubscript{1} amplitudes, however require more complex packaging than the above presented structure. 

\begin{figure*}[ht]
  \centering
  \includegraphics[width=\textwidth]{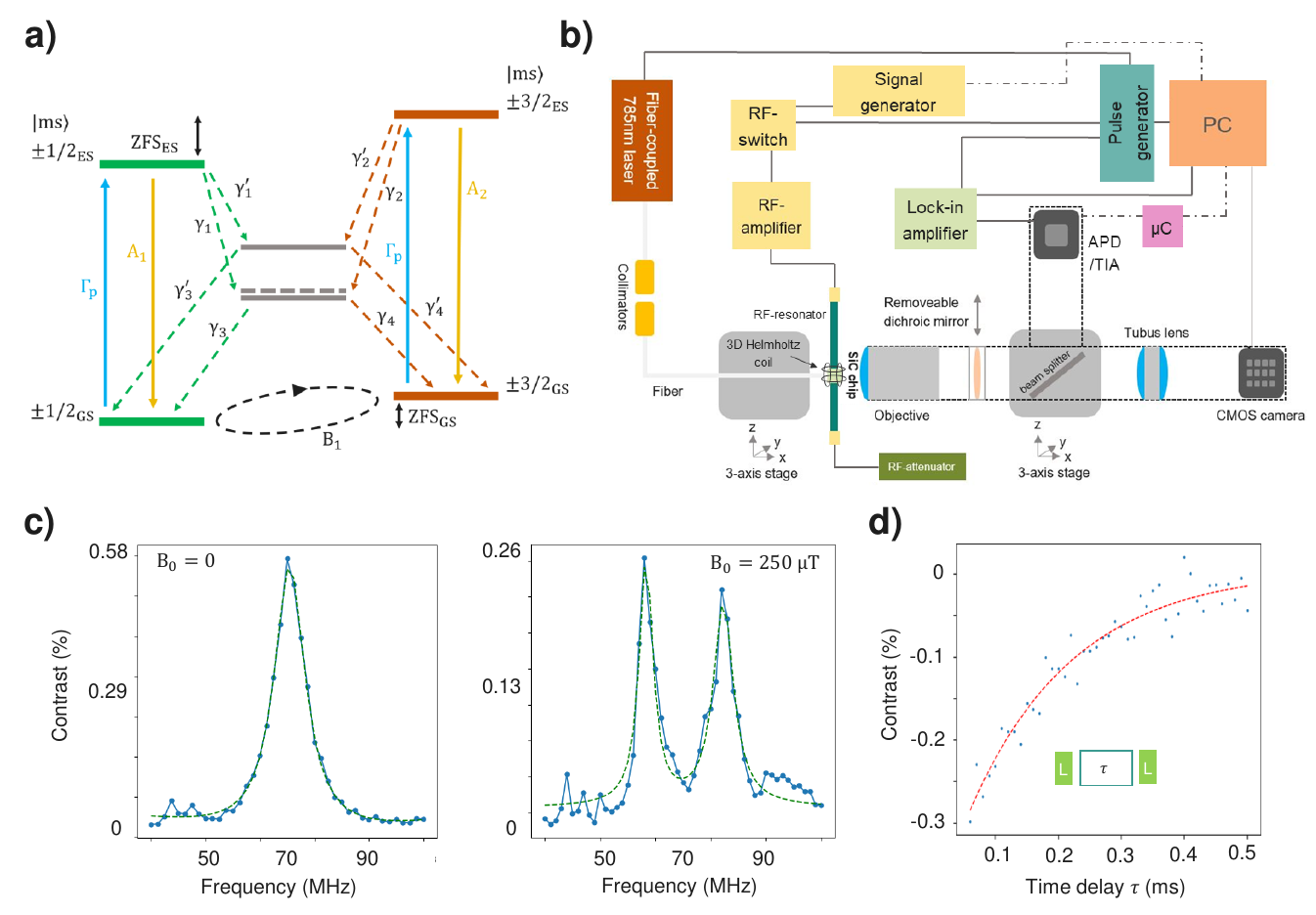}
  \caption{a) Hybrid quantum model for describing both the ground-state coherently as well as the fluorescence dynamics of the V2. b) Block diagram of characterization setup for both continuous wave and pulsed measurements. c) CW-ODMR measurements performed on the V2 ensemble embedded into the planar waveguide for the case of B\textsubscript{0}=0, and B\textsubscript{0}=250 $\upmu \mathrm{T}$, fitted with a single and double Lorentzian respectively. d) T1 measurement on the V2 ensemble in the waveguide by initializing the system with a first laser pulse and then reading out the fluorescence with a second laser pulse separated by a time-delay $\tau$.}
  \label{fig:Figure3}
\end{figure*}
\section{Results}
Based on proton and electron implantation techniques, we conducted a comparative study in order to identify the most effective method to generate V2 silicon vacancies. Proton implantation was performed on CVD-grown bulk epitaxies on n\textsubscript{++} 4H-SiC substrates with different implantation energies and doses, followed by annealing. The use of protons enables precise depth control, as they are completely stopped within the SiC bulk, with the applied energy determining the depth profile. Simulation results using the Stopping and Range of Ions in Matter (SRIM) model \cite{ziegler2012stopping} are presented in Figure \ref{fig:Figure2} c), illustrating the depth profiles for 400 keV and 600 keV protons, with peak depths of approximately 3 $\upmu$m and 5 $\upmu$m, respectively. Experimental verification of these results is provided in Figure \ref{fig:Figure2} b), which shows a chip irradiated with an energy of 400 keV, scanned from the side using a confocal setup with an immersion lens. The resulting depth profile of the generated V2 color centers reveals a peak located around 3 $\upmu$m from the bottom edge, which confirms the simulation results. Continuous-wave optically detected magnetic resonance (CW-ODMR) measurements were performed at the center of the implantation ribbon with confocal microscopy at around 0.7 mW laser power and around 22 dBm RF-power through an impedance-unmatched wire as in \cite{widmann2015coherent}. This measurement yields an ODMR contrast of approximately 1 percent between the resonance peak and background of the plot. This result confirms the successful generation of a dense V2 ensemble. By analyzing the count rate, a peak ensemble density of around 350 $\pm$ 80 / $\upmu \mathrm{m^3}$ color centers can be inferred. In addition to proton implantation, electron implantation was performed. On both proton- and electron-irradiated samples, low-temperature spectra were measured, revealing a distinct zero-phonon line (ZPL) at \SI{916}{nm}, as well as the phononic sideband still visible at 8 K, shown in Figure \ref{fig:Figure2} d). Due to the significantly smaller mass of electrons compared to protons, their comparatively long mean free path length allows them to pass through the entire wafer. It can thus be assumed that no implantation peak is present resulting in a homogeneous depth-distribution. This means that the strongest limiting factor is the local nitrogen doping for the electrical charge-state of the color centers \cite{widmann2019electrical}, where V2s are single-negatively charged. By correlating the N doping concentration (measured by standard capacitance-voltage measurement) with the ZPL fluorescence, we infer an optimal N-doping concentration of $10^{14}$-$10^{15}$/$\mathrm{cm^3}$.\\
\\
\begin{figure*}[ht]
  \centering
  \includegraphics[width=\textwidth]{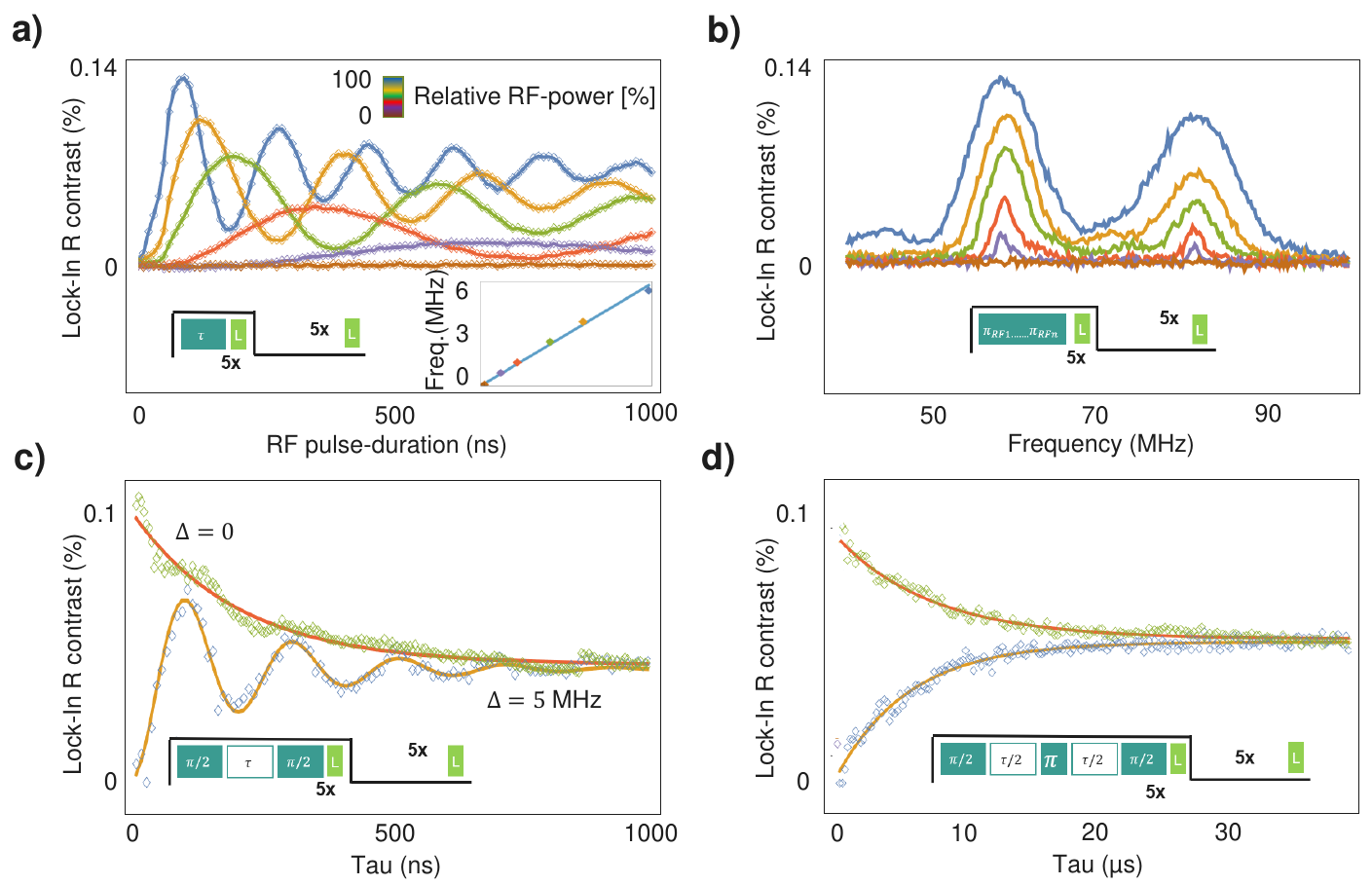}
  \caption{Pulsed measurements performed on the V2 ensemble in the planar waveguide, consisting of $\pi/2$ or $\pi$ RF-pulses, free evolution times $\tau$ and laser initialization and readout pulses $L$: a) Rabi-oscillations for different $B$\textsubscript{1} amplitudes fitted with exponentially decaying harmonic, b) Pulsed ODMR measurement for different amplitudes of $B$\textsubscript{1}, c) Ramsey sequence for a detuning of 0 and 5MHz fitted with a decaying exponential and an exponentially decaying harmonic for the case of detuning respectively, d) Hahn-echo sequence performed for two different phases of the $\pi/2$-pulse (x:top  and y:bottom) fitted with a decaying exponential.}
  \label{fig:Figure4}
\end{figure*}
The waveguide SiC chips were fabricated using a high‐volume 6‐inch wafer process. The SiC epitaxy was optimized based on the previous findings in terms of N-doping concentration and a layer of \SI{9}{\micro m} thickness was grown. Subsequently, a thin plasma-enhanced CVD oxide layer was deposited to form the upper waveguide cladding, resulting in a dual-mode waveguide \cite{stuermer2025low}. The N-doping in the epitaxy of each waveguide was chosen to have an optimized N-doping concentration at each peak where the ensembles are generated. Protons were implanted, chosing the ion-energy to make sure that the implantaton depth matches the maximum of the waveguide mode for optimal photonic coupling with the V2. Subsequent annealing was performed to remove residual crystal damage. Finally the chips were diced into 3x3 mm\textsuperscript{2} chips by an optimized process yielding low optical surface roughness for low photonic coupling losses \cite{stuermer2025low}. \\
\\
The V2 was modelled using the Lindblad equation \cite{singh2022multi}
\begin{equation} \label{eq:lindbladian}
\frac{d \rho}{dt} = -\frac i\hbar [H,\rho] + \sum_\mathrm{i} \left( L_\mathrm{i} \rho L_\mathrm{i}^\dagger -\frac 12 \{ \rho, L_\mathrm{i}^\dagger L_\mathrm{i}\} \right)
\end{equation}
where $\rho$ is the density matrix and $L_\textsubscript{i}, L_\textsubscript{i}^\dagger$ are an infinite set of jump operators acting on the Hilbert space. The mathematical form of the Lindblad interaction ensures that the trace of the density matrix stays constant. The full physical model is shown in Figure \ref{fig:Figure3} a). Here only the ground state is modelled (first 4 levels) as a coherent quantum system, described via the Von-Neumann equation and the excited and metastable states are modeled as a non-coherent rate-model, in total using a 10-dimensional Hilbert space to account for all levels. The form of the Lindblad jump operator for the diagonal states is \cite{singh2022multi}: 
\[
L_{\mathrm{i}} = \sqrt{R_\mathrm{i}} \ket {u_\mathrm{i}} \bra {v_\mathrm{i}}
\]
where $\ket {u_\textsubscript{i}}$ and $\ket {v_\textsubscript{i}}$ are the final and inital states of the transition and $R_\textsubscript{i}$ the transition rate (1/s), where the values are extracted from \cite{liu2024silicon}. Here, we note that those rates were measured at cryogenic temperatures and might deviate from room temperature. The Hamiltonian $H$ in equation \ref{eq:lindbladian} is the standard V2 4-level ground state Hamiltonian \cite{castelletto2023quantum}, taking into account both the applied magnetic field $B$\textsubscript{0} and $B$\textsubscript{1} driving fields. The spin-lattice relaxation and the dephasing of the ground state are modelled with the following Lindblad operators, taken from \cite{singh2022multi}:
\[
L_\mathrm{\alpha} = \sqrt{2 \alpha}\ S_\mathrm{x} \qquad L_\mathrm{\beta} = \sqrt{2\beta}\ S_\mathrm{z}
\]
The connection between those relaxation parameters and experimental quantities can be established by analyzing the corresponding dissipation matrices yielding $\alpha = \frac{1}{3 T_1}$ and $\beta = T_2^{-1} - \frac{5}{2} \alpha $. Combining $T_{1}$ data from Figure \ref{fig:Figure3} d) alongside with the extracted $T_{2}$ value in Figure \ref{fig:Figure4} d) we estimate $\alpha = 2123$ 1/s and $\beta = 3.54 \cdot 10^5$ 1/s.  \\
\\
We use a custom-built transmission fluorescence setup, which is depicted in Figure \ref{fig:Figure3} b). A 785 nm mode coming from the laser (pigtailed Hübner Photonics Cobolt 06-01) is coupled into a single-mode PM fiber which is then edge coupled into the SiC-waveguide containing V2 color centers via a 3-axis stage. On the opposite side, an objective (Nikon OFN25, 0.3 NA) was used to collect the emitted light. Next in the beam path, a dichroic mirror with cutoff wavelength of 850 nm (Thorlabs FELH0850) was used to filter out any unwanted excitation light. The V2 fluorescence light passes through a beamsplitter, where one arm gets focused onto a standard CMOS-camera (IDS uEye), while the other gets focused onto an avalanche photodiode (APD440A) with integrated transimpedance-amplifier (TIA) and variable gain. Due to the vertical photonic confinement but lateral free evolution \cite{stuermer2025low}, the CMOS camera shows a quasi-elliptical profile of the emitted light. The output voltage of the TIA is measured via a lock-in amplifier (Stanford Research Systems SR865A) (configuration 1, solid line) or a standard microcontroller (Arduino Uno Minima) (configuration 2, dashed line). On the transmitter side, we either control the signal generator (Rigol DG 2000) via serial connection with the PC for CW-measurements, which are not time-critical in terms of synchronization, or we use a programmable pulse-generator as in \cite{reisenbauer2022lithpulser} depending on the configuration. The pulse generator synchronizes the lock-in frequency, which demodulates the resulting voltage signal coming from the APD. In each pulse sequence, 5 identical sequences  are performed, due to the low bandwidth of the APD (APD440), requiring time-averaging. This would not be necessary with a high speed detector, fully capable of time resolving the fluorescence response. In the low phase of the rectangular modulation (compare Figure \ref{fig:Figure4}) just the laser is switched on, whereas in the high phase, also the RF-switch (Mini-Circuits ZASWA-2-50DR+) gets triggered, effectively modulating the signal. The RF-signal is amplified by a 43 dB amplifier (Mini-Circuits LZY-22+) and passes through the microstrip, before being dissipated in an attenuator (Microlab TB-80MN) in order to avoid RF-reflections.\\ 
\\
First, CW-ODMR measurements are performed for $B$\textsubscript{0}=\SI{0}{\upmu T} and $B$\textsubscript{0}=\SI{250}{\upmu T} for the \SI{9}{\upmu m} waveguide by use of a custom-built 3D-Helmholtz coil, applying the magnetic field parellel to symmetry axis as shown in Figure \ref{fig:Figure3} c) at a laser-power of 36 mW in the waveguide. To this end, the SiC chips are placed on a broadband PCB-based microstrip manufactured on Rogers 4003C substrate as shown in Figure \ref{fig:Figure3} b) with 3 mm width, which creates a very similar field structure as shown in the simulation, but with different amplitude. The resulting contrast without magnetic field is around 0.6 percent at zero field and \SI{0.26}{\%}  when a Zeeman splitting is induced. The shot-noise limited sensitivity for CW-ODMR is given \cite{barry2020sensitivity} by: 

\begin{equation} 
\eta_{\mathrm{cw}} = \frac{4}{3\sqrt{3}} \frac{h}{g_\mathrm{e} \mu_\mathrm{B}} \frac{\Delta v}{C_{\mathrm{cw}} \sqrt{R}}
\end{equation}
where $h$ is the Planck constant, $g_e$ the Land\'e-factor for the system, $\upmu_\mathrm{B}$ the Bohr magneton, $\Delta v$, the FWHM linewidth, $C_{\mathrm{CW}}$ the CW-ODMR contrast and $R$ the fluorescence photon number. We estimate a sensitivity below 270 nT/$\sqrt{\text{Hz}}$. In comparison the confocal CW-ODMR (Figure \ref{fig:Figure2}) delivers a sensitivity of around \SI{40}{\upmu T/\sqrt{\text{Hz}}}. \\
\\
Subsequently, pulsed measurements are performed. To this end, the microstrip is changed to a home-built coil with 10 windings and a 3D-printed frame, which equally creates a highly homogeneous magnetic field in the center of the coil but requires less current for the same $B$\textsubscript{1} compared to the sheet conductor in Figure \ref{fig:Figure2}. The laser initialization and readout times $t_I$ and $t_R$ are chosen to be \SI{12}{\upmu s}. Figure \ref{fig:Figure4} a) shows Rabi-oscillations using different $B$\textsubscript{1}-field amplitudes. Here we observe clear coherent oscillations in fluorescence, and thus population in the respective spin states, demonstrating that a large ensemble of an estimated $6.4 \cdot 10^7$ V2 centers can be driven coherently embedded in a planar waveguide. The Rabi-frequency increases linearly with the $B$\textsubscript{1} amplitude as expected theoretically. Next, pulsed ODMR measurements are performed, shown in Figure \ref{fig:Figure4} b) for different $B$\textsubscript{1} amplitudes. The shot-noise sensitivity \cite{barry2020sensitivity} is given by: 
\begin{equation} 
\eta_{\mathrm{pulsed}} = \frac{8}{3\sqrt{3}} \frac{\hbar}{g_\mathrm{e} \mu_\mathrm{B}} \frac{1}{C_{\mathrm{pulsed}} \sqrt{N}} \frac{\sqrt{t_\mathrm{I} + T_\mathrm{2}^* + t_\mathrm{R}}}{T_\mathrm{2}^*}
\end{equation}
where $N$ is the number of counts collected for each readout, and $T$\textsubscript{2}\textsuperscript{*} the decoherence time. Using this metric, a sensitivity of below 30 nT/$\sqrt{\text{Hz}}$ is estimated. Subsequently, the Ramsey protocol is tested successfully, seen in Figure \ref{fig:Figure4} c) for the case that the detuning between the microwave and the resonance $\Delta$ is 0 and 5 MHz, where the first case does not and the second case does show oscillations of 5 MHz as expected from theory \cite{barry2020sensitivity}. Moreover a $T$\textsubscript{2}\textsuperscript{*} coherence time of around 230 ns is extracted from the measurement. This might be improved in future since $B$\textsubscript{1} homogeneity was not optimized in this measurement, which causes effective dephasing across the emsemble. We estimate the sensitivity for the Ramsey protocol, which relies on measuring the phase of the resulting detuned oscillation in order to extract the B\textsubscript{0} amplitude, which is given by \cite{barry2020sensitivity}: 
\begin{equation} 
\eta_{\mathrm{Ramsey}} = \frac{\hbar}{\Delta m_\mathrm{s} g_\mathrm{e} \mu_\mathrm{B}} \frac{1}{C e^{-(\tau/T_\mathrm{2}^*)^\mathrm{p}} \sqrt{N}} \frac{\sqrt{t_\mathrm{I} + \tau + t_\mathrm{R}}}{\tau}
\end{equation}
where $\Delta m_s$ is the spin-difference between the two states and $p$=1 for Lorentzian peaks and $\tau$ is the free-precession time on the equator of the Bloch-sphere. A shot-noise limited sensitivity below 50 nT/$\sqrt{\text{Hz}}$ is estimated. Finally, Figure \ref{fig:Figure4} d) shows a demonstration of Hahn-echo for 2 different phases of the refocusing $\pi/2$ pulse. It gives a $T$\textsubscript{2} coherence time of around 2.8 $\upmu$s. In scenarios where one knows the frequency of AC-$B$\textsubscript{0} magnetic field to be measured, the Hahn-echo can be used to precisely measure it with the advantage of inherently suppressing low-frequency stray-fields, especially DC, which is invisible for Hahn-echo. Sensitivity is highest for fields with periods close to $\tau$ which is limited fundamentally by $T$\textsubscript{2}. In this case a $B$\textsubscript{0} field with around 300 kHz could be resolved with below 10 nT/$\sqrt{\text{Hz}}$ improving the Ramsey sensitivity by extending $T$\textsubscript{2}\textsuperscript{*} to $T$\textsubscript{2} \cite{barry2020sensitivity}. It shall be mentioned that those sensitivities are assuming shot-noise as the main sensor noise source and do not consider electronic noise sources in the laser and readout circuits. \\

\section{Conclusion}
This study presents a new approach that enhances and simplifies the sensitivity of silicon carbide (SiC)-based quantum sensors. We demonstrate an improvement in sensitivity, exceeding conventional confocal measurements by at least two orders of magnitude. The photonic concept employed is broadband, thereby offering the potential for expansion to other vacancy types. Furthermore, the underlying architecture of this approach enables a reduction in optical power consumption, rendering it a more energy-efficient solution compared to traditional methods and simplifying the sensor concept. Additionally, the SiC-based technology is fully volume-compatible, allowing for seamless integration into existing semiconductor manufacturing processes, in stark contrast to diamond-based systems which are often limited by their material properties and fabrication constraints. Despite the advancements achieved, there remains room for further improvement, such as optimizing the pulse sequences to achieve highest sensitivity, integrating the chip into a miniaturized module and further optimizing power density. This development may enable SiC-based quantum sensors to rival the performance of diamond-based systems in future and provides a clear path towards mass-production of quantum sensing technology.

\begin{acknowledgments}
This work was supported by funds of the BMFTR project QVOL
Grant No. 03ZU1110IA and European Union project 101189875 ACDCQ as well as by the Austrian Research Promotion Agency project FFG FO999914034 (SPQV). The authors would like to extend their sincere gratitude to the BMFTR funded QVOL consortium as well as the Austrian Academy of Sciences for their valuable contributions to the discussion. Special thanks shall be given to Yan Huck for very fruitful discussions and early contributions. 
\end{acknowledgments}

\nocite{*}
\bibliographystyle{apsrev4-2}
\bibliography{references}

\end{document}